\documentclass{raa}

\usepackage{graphicx,times}             

\begin{document}

   \title{Is the additional increase of the star luminosity due to partial mixing real?
}

   \volnopage{Vol.0 (20xx) No.0, 000--000}      
   \setcounter{page}{1}          

   \author{E. I. Staritsin
   }

   \institute{K.A. Barkhatova Kourovka Astronomical Observatory, B.N. Yeltsin Ural Federal University,
             pr. Lenina 51, Ekaterinburg 620000, Russia; {\it Evgeny.Staritsin@urfu.ru}\\
   }

   \date{Received~~2009 month day; accepted~~2009~~month day}

\abstract{ The partial mixing of matter between the radiative envelope and the convective core in the early type
B--star produces an additional increase of star luminosity during the main sequence evolution. The high quality
data on stellar mass and luminosity defined from the studies of detached double-lined eclipsing binaries
are used to check the existence of such additional increase. It is shown that the additional luminosity
increase does not contradict to the observed data of high quality, if the intensity of partial mixing is restricted
by the observed increase in surface helium content.
}

   \authorrunning{E. I. Staritsin }            
   \titlerunning{Is the additional increase of the star luminosity }  

   \maketitle

%
%
\section{Introduction}           
\label{sect:intro}

One of the observed properties of the early B main-sequence single stars is an increase in surface helium abundance with
stellar age. According to Lyubimkov et al.~(\cite{LRL04}), the change of number density ratio, $N_{He}/N_{H}$, reaches
26$\pm$10$\%$ in the surfaces of stars with masses 4$<M/M_\odot<$12 by the end of main sequence evolution. Here $N_H$
and $N_{He}$ are the number density of hydrogen and helium respectively. In stars with masses 12$<M/M_\odot<$19 this
change may be as large as 67$\%$. According to Huang and Gies~(\cite{HG06a}), such change among the stars with masses
8$<M/M_\odot<$16 is 23$\pm$13$\%$. The helium enrichment process acts more strongly in the faster rotators (Huang and Gies 2006a).
The increase in surface helium abundance among binary star components may be greater than among single stars as much
as twice (Lyubimkov et al.~\cite{LHRT95}, \cite{LHRT96}, \cite{LHRT97}; Tarasov et al.~\cite{THH95}).

The partial mixing of matter between the convective core and the radiative envelope in the rotating star may be a possible
mechanism of surface helium enrichment (Staritsin 2014, 2014a, 2017, 2017a). The rotating star is the site of hydrodynamic
processes of transport such as meridional circulation and shear turbulence. In the case of moderately rapid rotation the intensity
of shear turbulence is enough to remove some amount of hydrogen from the radiative envelope to the layer with variable chemical
composition. The semi-convective mixing begins to operate in that layer as a result of the hydrogen content increase. So hydrogen
flows from the radiative envelope through the semi-convective zone to the convective core. Helium flows in the opposite direction.
As a consequence some amount of helium removed from the convective core to the radiative envelope. The result is the surface helium
content increases. Another consequence is that some amount of hydrogen removed from the radiative envelope to the convective core.
This leads to both the synthesis of additional amount of helium and the additional increase of star luminosity during the main sequence
evolution (Fig.~1). Partial mixing of material in the star’s interior may influence on the mass-luminosity relation of main sequence
stars. Such influence is stronger for the faster rotators. In this paper we compare theoretical mass-luminosity relation with
observational one. To produce the observational relation we used a best quality data on stellar masses and radii defined for
the detached double-lined eclipsing binaries.

   \begin{figure}
   \centering
   \includegraphics[width=\textwidth, angle=0]{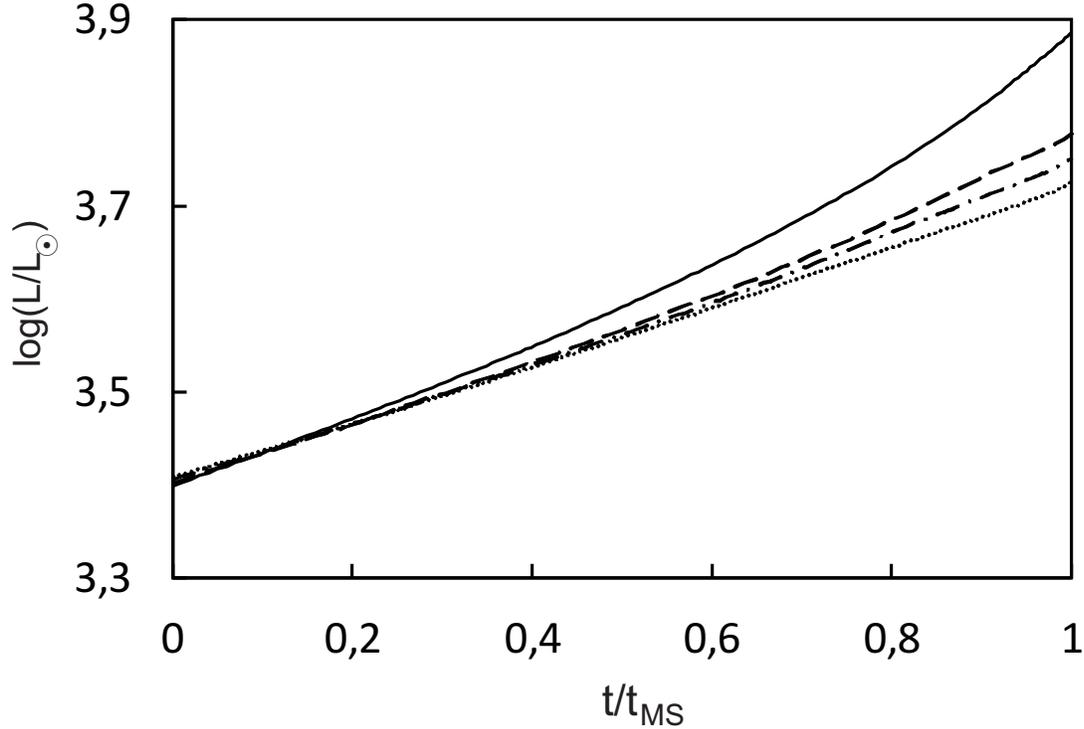}
   \caption{
   The luminosity of the star with mass 8$M_\odot$ as a function of the relative age $t/t_{MS}$
   for the initial equatorial rotational velocities $(V_e)_0$=250 km/s (solid), 200 km/s (dashed),
   150 km/s (dot-dashed), and 100 km/s (dotted). The data are taken from Staritsin (\cite{S17}). }
   \end{figure}

\section{Empirical mass-luminosity relation}

In order to get the empirical mass-radius and mass-luminosity relations a subset of the close binaries
components
with masses $7\le M/M_\odot\le23$ was picked out of the catalog of absolute elements collected by Malkov (\cite{M07}).
The data on mass, radius and luminosity of components placed in that catalog are based on the research of the detached
double-lined eclipsing binaries. There are 27 components with masses from the pointed interval of mass in the catalog.
Mass-radius and mass-luminosity relations are defined by a least squares method (Afifi, Azef~\cite{AA79}) as follow:
\begin{equation}
\lg R=(0.08\pm0.10)+(0.64\pm0.09)\lg M,
\label{eq:MR}
\end{equation}
\begin{equation}
\lg L=(0.94\pm0.18)+(3.01\pm0.17)\lg M,
\label{eq:ML}
\end{equation}
radius $R$, luminosity $L$, and mass $M$ are in solar units. The standard deviations are 0.07 for $\lg R$ and 0.11 for $\lg L$.



\section{Theoretical mass-luminosity relation}

\subsection{Spin Rotation of Binaries Components}
The spin velocity of the binaries components taken from the collected subset lies in between
of 20 km/s and 180 km/s. The mean value of spin velocity of that
components subset is equal to 118 km/s. The single stars lying in the same mass interval,
for which helium enhancement was observed, show mean value of rotational velocity of 174 km/s
near ZAMS and of 134 km/s near TAMS (Huang, Gies~\cite{HG06}). So the subset of binaries
components is characterized by slower rotation.

\subsection{Mixing Processes in Binary Components Interiors}
As in the case of single star the spin rotation of the binary star component is the reason
of the meridional circulation and shear turbulence. But these mixing processes are not too active
because of the lower velocity of binary component spin rotation. Apart from the rotational disturbance
the binary components are influenced by the mutual tidal interaction. The tidal interaction may give
rise to both thermally (Tassoul J., Tassoul M.~\cite{TT82}) and mechanically (Tassoul J.~\cite{T87};
Tassoul J., Tassoul M.~\cite{TT90}) driven circulation. Power of mixing processes
generated by the rotational disturbance and the tidal interaction are characterized by the centrifugal
acceleration and the tidal acceleration respectively. The ratio of centrifugal acceleration of the
component $a_{spin}$ to tidal one $a_{tidal}$
\begin{equation}
\frac{a_{spin}}{a_{tidal}}=\frac{M+M_{sec}}{2M_{sec}}\left(\frac{P_{orb}}{P_{spin}}\right)^2
\label{eq:AR}
\end{equation}
does not much differ from unity in the binary star composed from stars of approximately equal mass
in the case of synchronization of spin and orbital rotation. Here $M$ and $P_{spin}$ -- mass of the component
and period of its spin rotation, $M_{sec}$ -- mass of the second star, $P_{orb}$ - orbital period of binary
system. The tidal interaction may enlarge the intensity of partial mixing in the binary components
interiors as much as twice. The slowly rotating component of a binary star may be mixed as much as a single star
in the case of moderately rapid rotation.

\subsection{Evolution of Binary Components with Partial Mixing}
Mixing induced by spin rotation may be reduced to one dimensional problem (Zahn~\cite{Zahn92}). This is not the case of
tidal induced mixing. The hydrogen burning stage of the binary components with masses of 6$M_\odot$, 8$M_\odot$,
12$M_\odot$, 16$M_\odot$, 20$M_\odot$, and 24$M_\odot$ was calculated using the parametrical description of partial mixing
of matter between the convective core and the radiative envelope of the star (Staritsin~\cite{S18}). The change of number
density ratio by the end of main sequence evolution, $\triangle(N_{He}/N_{H})$, was chosen as a parameter.
The possible values of that parameter are restricted
by the observed increase in surface helium content (Huang, Gies~\cite{HG06a}; Lyubimkov et al.~\cite{LHRT95},
\cite{LHRT96}, \cite{LHRT97}, \cite{LRL04}; Tarasov et al.~\cite{THH95}). The usual overshooting parameter $\alpha$
was applied to determine position of the convective core boundary. This parameter is restricted by the observed
main sequence width of open clusters (Schaller et al.~\cite{SSMM92}; Meynet et al.~\cite{MMM93}). There were
calculated five variants of stellar evolution: $(\triangle(N_{He}/N_{H}),\alpha)=\{(0\%,0.05),(40\%,0.05),(80\%,0.05),
(0\%,0.15),(0\%,0.25)\}$. Calculated variants of stellar evolution encompass the helium enlargement of radiative
envelopes produced by the hydrodynamic calculation of single stars evolution (Staritsin~\cite{S17}).

\subsection{Mean relative age of Binary Components Subset}
Some evolutionary sequences and all binaries components taken from the selected subset are shown on Fig.~2.
It is obvious that the stars staying on late stage of hydrogen burning are absent in the selected subset
of binaries components. For each calculated variant of stellar evolution we determine the relative age $t/t_{MS}$
of each binaries component using its mass and radius. Here $t$ is the age of the component, and $t_{MS}$ is
its time of life on hydrogen burning stage. Then we calculate the mean value of relative age
of the components subset for each variant of stellar evolution. The mean values of relative age are
0.40, 0.43, 0.47, 0.38 and 0.37 for considered variants of stellar evolution respectively.

   \begin{figure}
   \centering
   \includegraphics[width=\textwidth, angle=0]{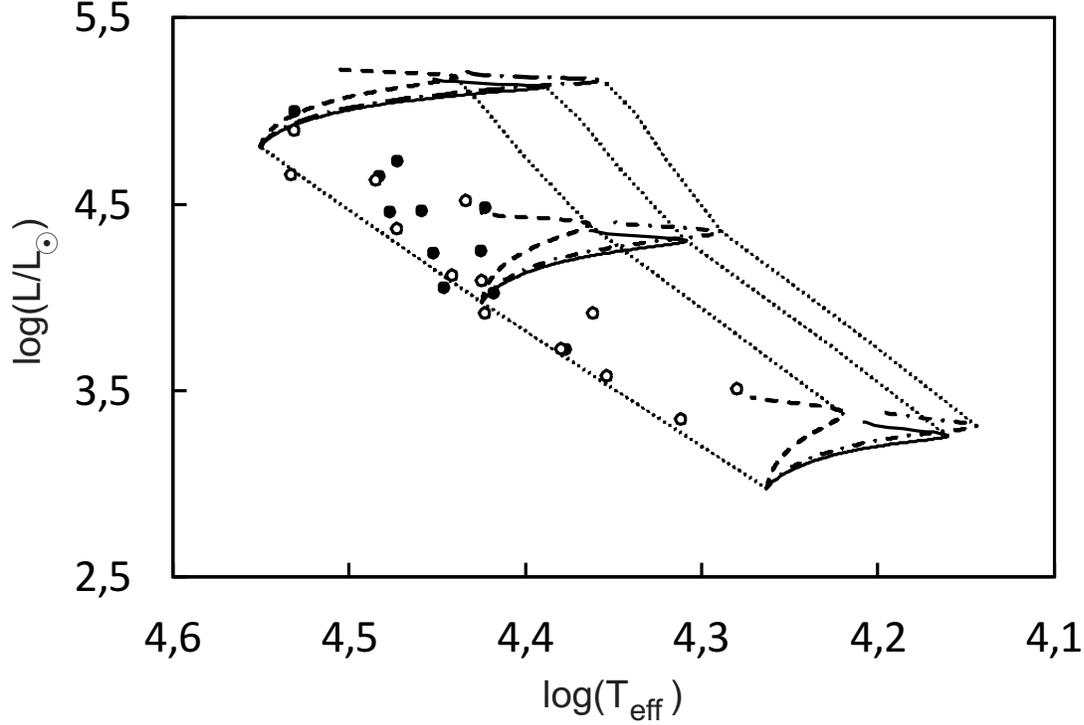}
   \caption{
   The evolutionary tracks for the stars with masses 6$M_\odot$, 12$M_\odot$, and 24$M_\odot$
   for the variants ($\triangle(N_{He}/N_{H})$, $\alpha$)=(0, 0.05) (solid), (40$\%$, 0.05) (dashed),
   and (0, 0.15) (dot-dashed). ZAMS and TAMSs are also shown (dotted). Filled circles denote
   primary components, open circles –- secondary components. The data on tracks are taken from
   Staritsin (\cite{S18}). The data on component elements are taken from Malkov (\cite{M07}). }
   \end{figure}

The theoretical mass-radius and mass-luminosity relations are built for the mean relative age of binary
components subset (Fig.~3). For each
evolutionary track
the mean value of luminosity and the mean value of radius are determined according to formula:
\begin{equation}
\overline{\lg\xi}=\sum_{j=1}^{10} w_j\overline{\lg\xi_j},
\label{eq:MV}
\end{equation}
$\xi=L$ and $\xi=R$ respectively. $\overline{\lg\xi_j}$ -- the mean value calculated using evolutionary
sequences over interval:
\begin{equation}
0.1(j-1)\le t/t_{MS}\le0.1j.
\label{eq:In}
\end{equation}
$w_j$ -- the relative number of components caught into the interval~(5). Due to applied method the theoretical
mass-radius relations for different variants of evolution do not much differ from the observed relation~(1)
and from each other (Staritsin~\cite{S18}).

\section{Discussion}
The theoretical mass-luminosity relation is influenced by the partial mixing of matter between the radiative
envelope and the convective core of the star (Fig.~3a).The deviation of the theoretical mass-luminosity
relations from the observed relation is no more than one standard deviation. This is because of both the number
of binary star components with known absolute elements is not too many and the mean value of relative age of those
components is not too much. The additional luminosity increase of stellar models produced by the evolution with partial
mixing of matter between the radiative envelope and the convective core of the star does not contradict to the observed
data of high quality.

   \begin{figure}
   \centering
   \includegraphics[width=\textwidth, angle=0]{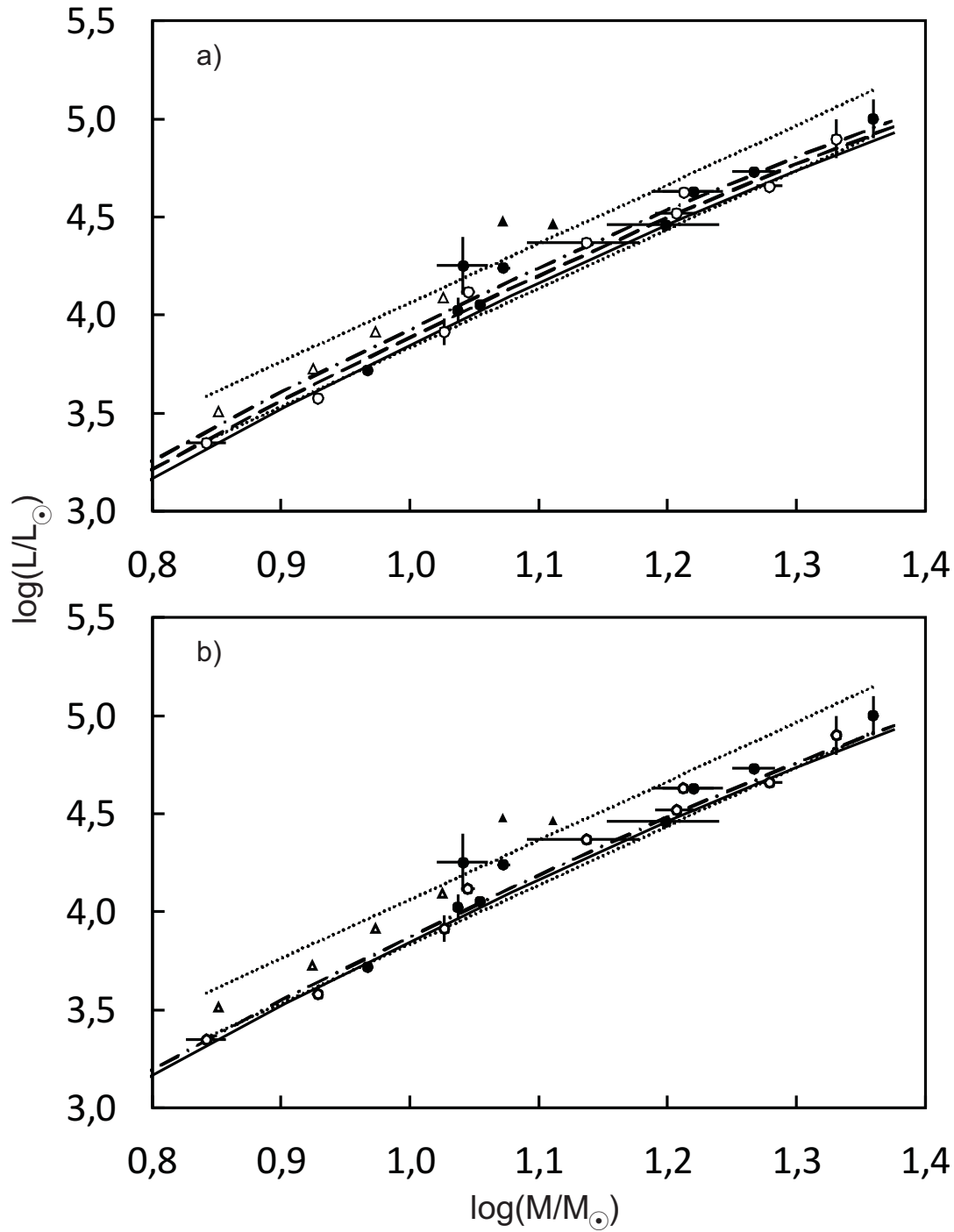}
   \caption{
   The mass-luminosity relation for the variants a) ($\triangle(N_{He}/N_{H})$, $\alpha$)=(0, 0.05) (solid),
   (40$\%$, 0.05) (dashed), (80$\%$, 0.05) (dot-dashed), b) (0, 0.05) (solid), (0, 0.25) (dot-dashed).
   Filled badges denote primary components, open badges –- secondary components. Observed error bars are shown
   in cases when the error bar surpasses the size of badge. Triangles denote cases when the error bar is unknown.
   Up and down standard deviations from empirical relation (2) are shown by dotted lines. The data on component
   elements are taken from Malkov (\cite{M07}). }
   \end{figure}

The theoretical mass-luminosity relation is practically not influenced by the additional mixing at the convective
core boundary (Fig.~3b). This is because of the additional luminosity increase got by a star by the mean relative
age of components subset depends only moderately on the overshooting
parameter $\alpha$. The overshooting parameter should be determined using the observed width of main sequence
built on the base of data on absolute elements of binaries components (Popova, Tutukov~\cite{PT90}).

\section{Conclusions}
\label{sect:conclusion}
The partial mixing of matter between the radiative envelope and the convective core of the single star introduced
in Staritsin~(\cite{S14}, \cite{S14a}, \cite{S17}, \cite{S17a}) to explain the observed increase in surface helium
content in the main sequence early type B--stars gives rise to the additional luminosity increase during the hydrogen
burning stage. The additional luminosity increase grows with time of star evolution, and it is larger among rapid
rotators. The components of binary systems with known absolute elements are characterized by lower spin rotation.
However there are the additional mixing processes in component interior produced by tidal interaction.
Such additional mixing processes may enlarge the intensity of partial mixing in the binary star components.
The mean age of the components with known absolute elements approximately equals to the half of main sequence life time.
The additional luminosity increase produced by that moment doesn't contradict to the observed luminosity
of binary components. This conclusion doesn't depend on the core overshooting, if the overshooting parameter
is restricted by the observed main sequence width of open clusters.  The most additional increase of luminosity
is produced by the end of main sequence evolution. Unfortunately the subset of components with known absolute elements
doesn't contain any information about the latest stage of core hydrogen burning.

\begin{acknowledgements}
This work was supported in part by the Ministry of Education and Science (the basic part of the State
assignment, RK no. AAAA-A17-117030310283-7) and by the Act no. 211 of the Government of the Russian
Federation, agreement no. 02.A03.21.0006.
\end{acknowledgements}

%

\label{lastpage}

\end{document}